\documentclass{aa}  

\usepackage{graphicx}
\usepackage{txfonts}
\usepackage{natbib}
\bibpunct{(}{)}{;}{a}{}{,} 

\begin{document}

   \title{RadioAstron as a target and as an instrument: Enhancing the Space VLBI mission's scientific output}


      \author{
          D.A.~Duev
          \inst{1,2}
          \and
          M.V.~Zakhvatkin
          \inst{3}
          \and
          V.A.~Stepanyants
          \inst{3}
          \and
          G.~Molera Calv\'{e}s
          \inst{1,4}
          \and
          S.V.~Pogrebenko
          \inst{1}
          \and
          L.I.~Gurvits
          \inst{1,5}
          \and
          G.~Cim\`{o}
          \inst{1}
          \and
          T.M.~Bocanegra~Baham\'{o}n
          \inst{1,5,6}
          }
   \institute{Joint Institute for VLBI in Europe, P.O. Box 2, 7990 AA Dwingeloo,
              The Netherlands\\
              \email{duev@jive.nl, molera@jive.nl, pogrebenko@jive.nl, lgurvits@jive.nl, cimo@jive.nl,
                         bocanegra@jive.nl}
         \and
             Sternberg Astronomical Institute, Lomonosov Moscow State University,
             Universitetsky av. 13, 119991 Moscow, Russia
        \and
            Keldysh Institute for Applied Mathematics, Russian Academy of Sciences,
            Miusskaya sq. 4, 125047 Moscow, Russia\\
	    \email{zakhvatkin@kiam1.rssi.ru, vic-stepan@rambler.ru}
       \and 
             Mets\"ahovi Radio Observatory, Aalto University, 02540 Kylm\"al\"a, Finland
        \and
            Department of Astrodynamics and Space Missions,
            Delft University of Technology, 2629 HS Delft, The Netherlands
        \and
            Shanghai Astronomical Observatory,
            80 Nandan Road, Shanghai 200030, China
             }

   \date{Received 8 September 2014 / accepted 27 October 2014}

 
  \abstract
   {The accuracy of orbit determination has a strong impact on the scientific output of the Space VLBI mission RadioAstron.}
   {The aim of this work is to improve the RadioAstron orbit reconstruction by means of sophisticated dynamical modelling of its motion in combination with multi-station Doppler tracking of the RadioAstron spacecraft.}
   {The improved orbital solution is demonstrated using Doppler measurements of the RadioAstron downlink signal and by correlating VLBI observations made by RadioAstron with ground-based telescopes using the enhanced orbit determination data.}
   {Orbit determination accuracy has been significantly improved from $\sim600$ m in 3D position and $\sim2$ cm/s in 3D velocity to several tens of metres and mm/s, respectively.}
   {}

   \keywords{astrometry --
                techniques: interferometric --
                instrumentation: interferometers --
                instrumentation: miscellaneous
               }

   \maketitle
%

\section{Introduction}

The Russian Federal Space Agency's RadioAstron mission uses a ten-metre parabolic antenna on board a dedicated Spektr-R spacecraft, which was launched into a highly elliptical orbit in July 2011 to perform observations of galactic and extragalactic radio sources as a space-borne component of a Space Very Long Baseline Interferometry (SVLBI) system \citep{2013ARep...57..153K}. 

Accurate knowledge of the spacecraft position is paramount for successful SVLBI observations. The mission's nominal orbit determination (OD) accuracy provided by the Ballistic Centre of the Keldysh Institute of Applied Mathematics of the Russian Academy of Sciences (KIAM) is at the level of $\sim$ 600 metres in 3D position and $\sim$ 2 cm/s in 3D velocity for the final orbit model available two to three weeks after any given date. This level of accuracy does allow SVLBI response (so-called fringes) to be detected in many cases, but is a limiting factor for advanced scientific applications, in particular, because of the unknown acceleration of the spacecraft.

The main factors determining the complicated orbital dynamics of RadioAstron include:

\begin{enumerate}
\item The necessity of regularly (about once a day) unloading the reaction wheels by firing thrusters, which perturbs the orbit. The reaction wheels experience a high load due to solar radiation pressure (SRP) exerted on the spacecraft structure, especially the 10 m radio telescope's antenna;

\item The orbit of the spacecraft being perturbed by the Moon. It was designed this way to allow various parts of the sky to be observed on a range of baselines during the mission lifetime, but at the cost of additional complications in orbit determination.
\end{enumerate}

High-accuracy orbit determination opens new scientific opportunities for RadioAstron. In particular, the current experience with RadioAstron suggests that its performance at the highest observing frequency of 22 GHz is limited by the orbit determination accuracy. Precise spacecraft state-vector determination will ensure the high sensitivity of RadioAstron necessary for tasks such as VLBI polarimetry at 22 GHz. SVLBI polarimetry at this frequency offers the potential for probing magnetic field structure in extragalactic relativistic jets on spatial scales never probed before.

In RadioAstron OD, data obtained from ground-tracking stations and on-board measurements are utilised. Routine tracking is performed by two Russian stations -- near Ussuriysk with a 70-m antenna and at Bear Lakes with a 64-m antenna. The data usually contain two-way range and Doppler observations (uplink frequency -- 5.7 GHz, downlink -- 3.4 GHz). The RadioAstron spacecraft is equipped with an H-maser frequency standard, which supports VLBI data acquisition and provides stable heterodyne reference for 8.4 GHz and 15 GHz downlink signals. The latter carries down the encoded observational data of the space radio telescope. During observations of a radio source, both signals are transmitted to a receiving station on Earth, either in Pushchino, Russia (station code Pu), or in Green Bank, West Virginia, USA (station code Gn). In the nominal regimes of RA operation that are used most often, a station observes both 8.4 and 15 GHz signals thereby enabling estimates of one-way Doppler. The other nominal regime when RA operates in a two-way mode implies no use of the on-board H-maser. The spacecraft also features an array of retroreflectors suitable for laser tracking. However, laser tracking is performed only occasionally, since among other conditions, it requires certain and not always attainable attitude of the spacecraft and favourable weather conditions. In addition, optical observations of RA spacecraft provide another OD data source.

On-board measurements generally feature attitude quaternions, rotation speeds of the reaction wheels, and operation parameters of thruster firings during an unloading, such as times of execution, duration, and mass of spent propellant. Attitude data are the most important since the direction of the change in spacecraft's velocity due to each unloading and SRP depend on it. Most of the time SRP is the only source of perturbing torque. During a radio astronomy science session, the spacecraft keeps constant attitude with respect to an inertial frame, therefore the angular momentum of reaction wheels only changes because of the perturbing torque, which, in turn, can be estimated if the rotation speeds of reaction wheels are known. Thrusters' operation parameters help to estimate the magnitude of velocity change $\Delta\mathbf{v}$ thanks to an unloading of reaction wheels. The spacecraft is not equipped with accelerometers so that non-gravitational perturbations due to SRP and unloadings cannot be measured directly. 

In an attempt to help improve the OD, Doppler observations with the European VLBI Network (EVN) telescopes of the RadioAstron spacecraft via its 8.4 GHz downlink signal have been conducted using the technique of Planetary Radio Interferometry and Doppler Experiments (PRIDE) led by the Joint Institute for VLBI in Europe (JIVE) \citep{2012A&A...541A..43D}. PRIDE is designed as a multi-purpose, multi-disciplinary enhancement of space missions science return that is able to provide ultra-precise estimates of spacecraft state vectors based on the phase-referenced VLBI tracking and radial Doppler measurements. These can be used for a variety of scientific applications (see, e.g., \citet{2014A&A...564A...4M}).

In this work, a new dynamical model of the RadioAstron motion and a new technique for additional utilisation of on-board measurements have been developed at KIAM. We demonstrate the quality of the improved orbital solution using PRIDE Doppler measurements of RadioAstron's downlink signal and the impact on data processing of RadioAstron observing science data.


\section{Dynamical modelling of RadioAstron spacecraft motion for orbit determination and orbit estimation pipeline}
The RadioAstron mission is not an easy one for orbit determination, although OD is an important ingredient in SVLBI. For RadioAstron, the development of a consistent dynamical model of the spacecraft is associated with a number of challenges. The most important one is defined by a significant amount of solar radiation pressure exerted on the spacecraft's surface, which produces both a perturbing acceleration and a torque. The acceleration, which has an average value of about $1.5\cdot 10^{-7}$ m/s$^2$, strongly depends on the spacecraft attitude and, owing to a complicated spacecraft's surface, has non-zero components in the direction, orthogonal to the Sun. Spacecraft dynamics are also complicated by the side effects of the stabilisation system's operation. The spacecraft maintains constant attitude with respect to an inertial frame except for the times when the orientation is changed intentionally to, for example, repoint the spacecraft to another radio source or to cool down the receiver while not observing. Attitude stabilisation is performed by means of reaction wheels. External torque leads to gradual build-up of angular momentum of reaction wheels and its subsequent unloading via stabilisation thrusters. Unloadings generally occur one to two times per day and produce a change in velocity of the spacecraft centre of mass of 3--5 mm/s in magnitude at a time. Both perturbations due to SRP and to unloadings of reaction wheels can be a source of significant orbital inaccuracy and require proper modelling.

The newly developed dynamical model of RadioAstron spacecraft motion is described in Appendix \ref{app1}. The orbit estimation pipeline is presented in Appendix \ref{app2}.

\section{Testing the improved orbit}

The approach described in Appendices \ref{app1} and \ref{app2} was applied to an orbital arc of RadioAstron spanning from February 20 to April 10, 2013. In the middle of this interval, a number of SVLBI observations were carried out, providing a test case for the orbital solution quality. 

The joint RadioAstron-EVN experiment GK047A was aimed at full-polarisation SVLBI imaging of the high-redshift quasar \object{0642+449} at 1.6 GHz as a part of the RadioAstron early science programme. These observations were conducted from 10:00 UTC, 2013.03.09 to 01:00 UTC, 2013.03.10. Participating telescopes and observational time ranges for particular telescopes are summarised in Table~\ref{gk047}. Although the maximum projected baseline was about 1 Earth diameter, the geocentric distance of RadioAstron during this experiment ranged from 120,000 to 190,000 km, which is more important for the purposes of this test.

\begin{table*}
\caption{Experiment GK047. Participating stations and observational UTC time ranges.}             
\label{gk047}      
\centering                          
\begin{tabular}{l c c c c}
\hline\hline                 
Antenna			&  Code	&	Country & GK047A & GK047B \\
\hline                        
Badary 			&	Bd	&	RU	&	20:00 -- 01:00 	&   --	\\
Effelsberg 			&	Ef	&	DE	&	10:00 -- 01:00 	&   --	\\
Evpatoria 			&	Ev	&	UA	&	10:00 -- 01:00 	&   --	\\
Greenbank		&	Gb	&	US	&	10:00 -- 01:00 	&   --	\\
Hartebeesthoek		&	Hh	&	ZA	&	10:00 -- 01:00 	&   --	\\
Jodrell Bank (Lowell)	&	Jb1	&	UK	&	10:00 -- 01:00 	&   --	\\
Medicina			&	Mc	&	IT	&	-- 	&   12:00 -- 01:00	\\
Mets\"ahovi		&	Mh	&	FI	&	-- 	&   10:00 -- 01:00	\\
Noto				&	Nt	&	IT	&	10:00 -- 01:00 	&   --	\\
Onsala60			&	On20	&	SE	&	-- 	&   10:00 -- 01:00	\\
Onsala85			&	On25	&	SE	&	10:00 -- 01:00 	&   --	\\
Robledo			&	Ro	&	ES	&	-- 	&   --	\\
Sheshan			&	Sh	&	CN	&	10:00 -- 01:00 	&   --	\\
Svetloe 			&	Sv	&	RU	&	-- 	&   20:00 -- 01:00	\\
Torun			&	Tr	&	PL	&	10:00 -- 01:00 	&   --	\\
Urumqi			&	Ur	&	CN	&	10:00 -- 01:00 	&   --	\\
Westerbork		&	Wb	&	NL	&	10:00 -- 01:00 	&   --	\\
Wettzell			&	Wz	&	DE	&	-- 	&   10:00 -- 01:00	\\
Zelenchukskaya		&	Zc	&	RU	&	20:00 -- 01:00 	&   --	\\
\hline
\end{tabular}
\end{table*}

The experiment GK047A was accompanied by the simultaneous PRIDE observations at 8.4 GHz with 5 EVN stations of the RadioAstron downlink telemetry signal designated with the code GK047B. Participated stations and their observational time ranges are given in Table~\ref{gk047}.

The orbital solution was put to the test using two independent approaches. First, the topocentric Doppler detections of the GK047B experiment were checked against model values, calculated using the improved orbit versus the values based on using the nominal orbital solution. Second, the VLBI data acquired by RadioAstron during the GK047A observational run were correlated with those of the ground-based telescopes. Similarly, two sets of model VLBI delays for RadioAstron were used, based on the nominal and the improved orbits.

\subsection{Doppler observations of the RadioAstron downlink signal}

For the GK047B experiment, the narrow-band Doppler data processing and carrier frequency extraction were conducted at JIVE with the SWSpec/SCtracker/DPLL software developed at the Mets\"{a}hovi Radio Observatory, Finland, in collaboration with JIVE\footnote{Wagner,~J., Molera Calv\'{e}s,~G. and Pogrebenko,~S.V. 2009-2014, Mets\"{a}hovi Software Spectrometer and Spacecraft Tracking tools, Software Release, GNU GPL, \url{http://www.metsahovi.fi/en/vlbi/spec/index}} \citep{GuifrePhD}. The output sampling time of our Doppler detections is ten seconds.


   \begin{figure}
   \centering
   \includegraphics[width=260pt,clip]{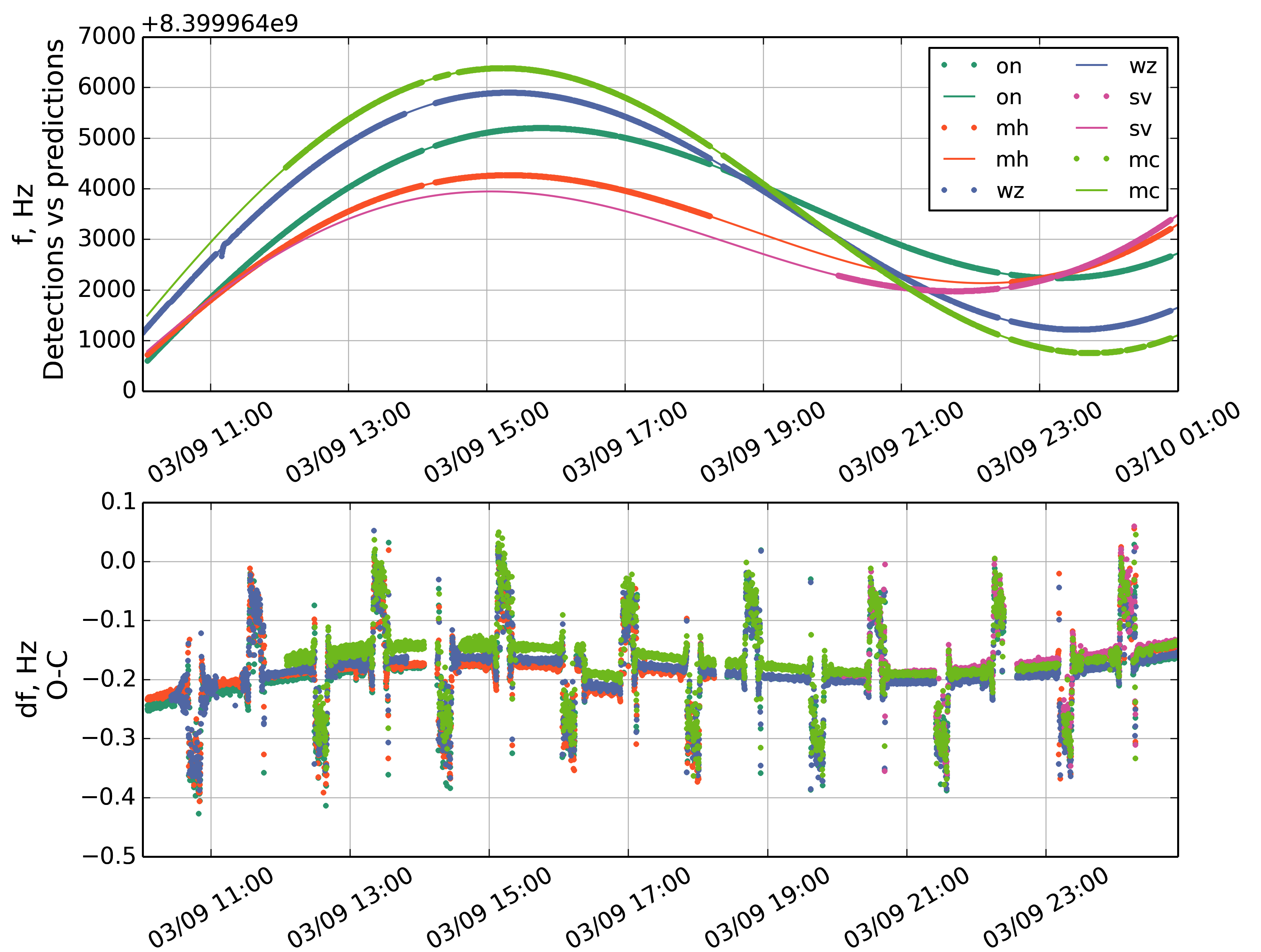}
   \caption{({\bf top}) Observed (O) topocentric RadioAstron downlink signal carrier frequencies (thick lines) versus computed (C) model values (thin lines). ({\bf bottom}) Differential frequencies (observed minus computed using the nominal orbital solution). Predictions account for geometrical, instrumental (including clock offsets/rates with respect to the GPS time scale), and propagation effects. Experiment GK047A, 2013.03.09-10.}
   \label{OCOldOrbit}
    \end{figure}

The detected topocentric frequencies of the RadioAstron's downlink signal carrier were checked against the computed model values (see Fig. \ref{OCOldOrbit}, top). The one-way topocentric Doppler predictions were calculated using the software package PYP (PYthon tools for PRIDE). This package coded in python programming language is based on the VINT software \citep{2012A&A...541A..43D}, which was considerably re-engineered for the JIVE's operational purposes. 
The theoretical model used for calculating predictions is described in \citet{duevPhD}. 

The differential frequencies (observed minus computed) demonstrate the quality of the orbital solution used for calculating the model values (see Fig. \ref{OCOldOrbit}, top)\footnote{Owing to a wide dynamic range any difference between the predictions based on nominal and improved orbits is not seen}. In the case of the nominal orbital solution, a large offset and a significant drift of the differential frequencies are clearly seen (Fig. \ref{OCOldOrbit}, bottom). We also note large outliers at the beginning and at the end of each scan on frequency plots that result from the RadioAstron's transmitter requiring cooling in between the scans.


   \begin{figure}
   \centering
   \includegraphics[width=270pt,clip]{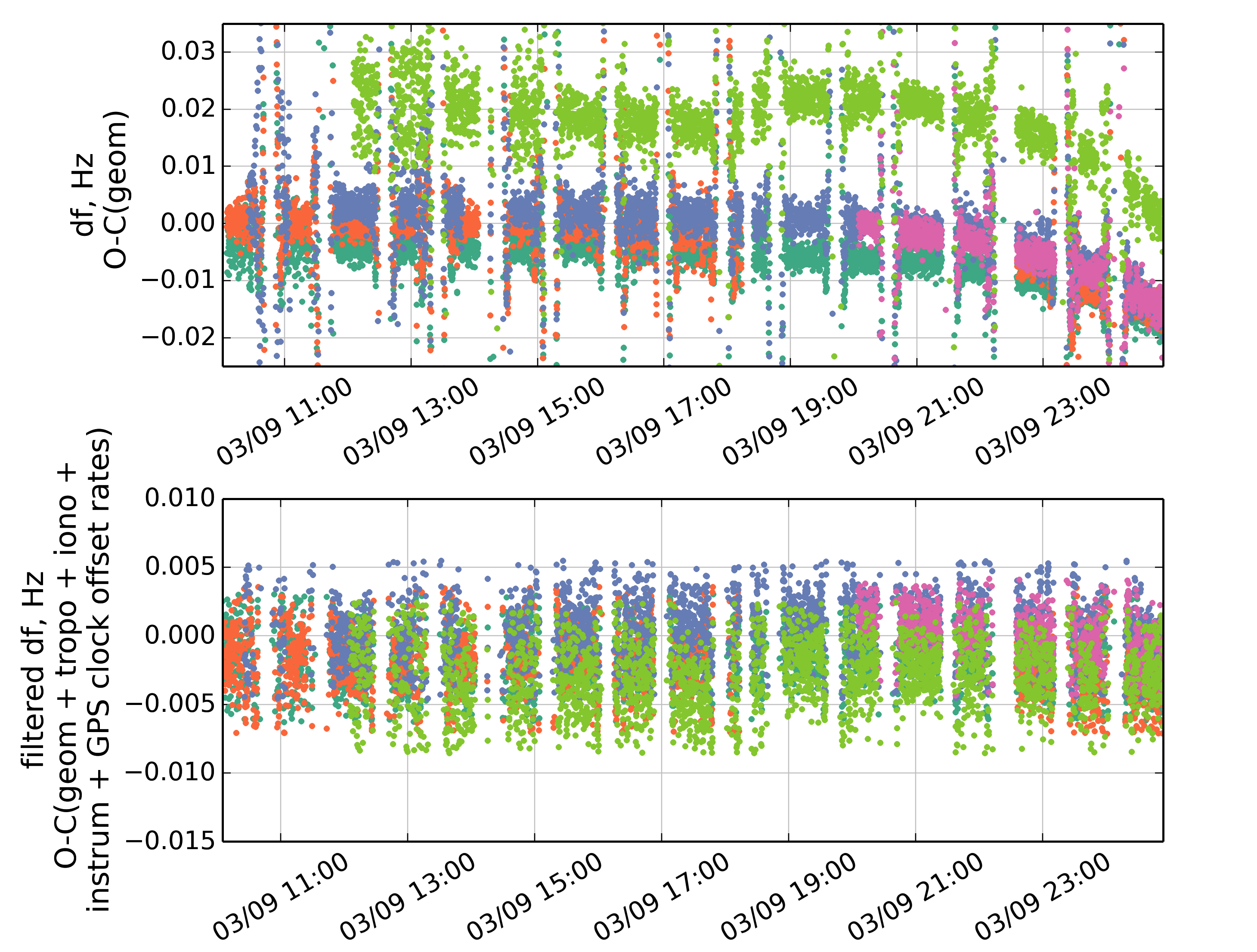}
   \caption{({\bf top}) Differential frequencies (observed minus computed). Predictions only take geometrical effects into account. ({\bf bottom}) Differential frequencies (observed minus computed). Predictions account for geometrical, instrumental (including clock offsets/rates with respect to the GPS time scale), and propagation effects. Epochs when RadioAstron's transmitter was cooling are filtered out.
Station colour codes used correspond to those from Fig. \ref{OCOldOrbit}.
Experiment GK047B, 2013.03.09-10.}
   \label{OCNewOrbit}
    \end{figure}

We then studied the differential frequencies for the improved orbital solution. Two cases were investigated: when predictions only account for the geometrical effects in a general relativistic sense, taking all sorts of geophysical effects into account in compliance with the IERS Conventions 2010 \citep{tn36} and when the instrumental (including clock offsets/rates at the stations with respect to the GPS time scale) and propagation effects are considered as well. These are shown in Fig. \ref{OCNewOrbit} (top and bottom), respectively.
The outliers mentioned above can reach $\sim 0.2$ Hz in magnitude and are therefore not seen in Fig. \ref{OCNewOrbit} (top), which emphasises the impact of instrumental and propagation effects on the detected frequency. These outliers were filtered out in Fig. \ref{OCNewOrbit} (bottom) using a simple $3\sigma$-criterion.

%
   \begin{figure}
   \centering
   \includegraphics[width=260pt,clip]{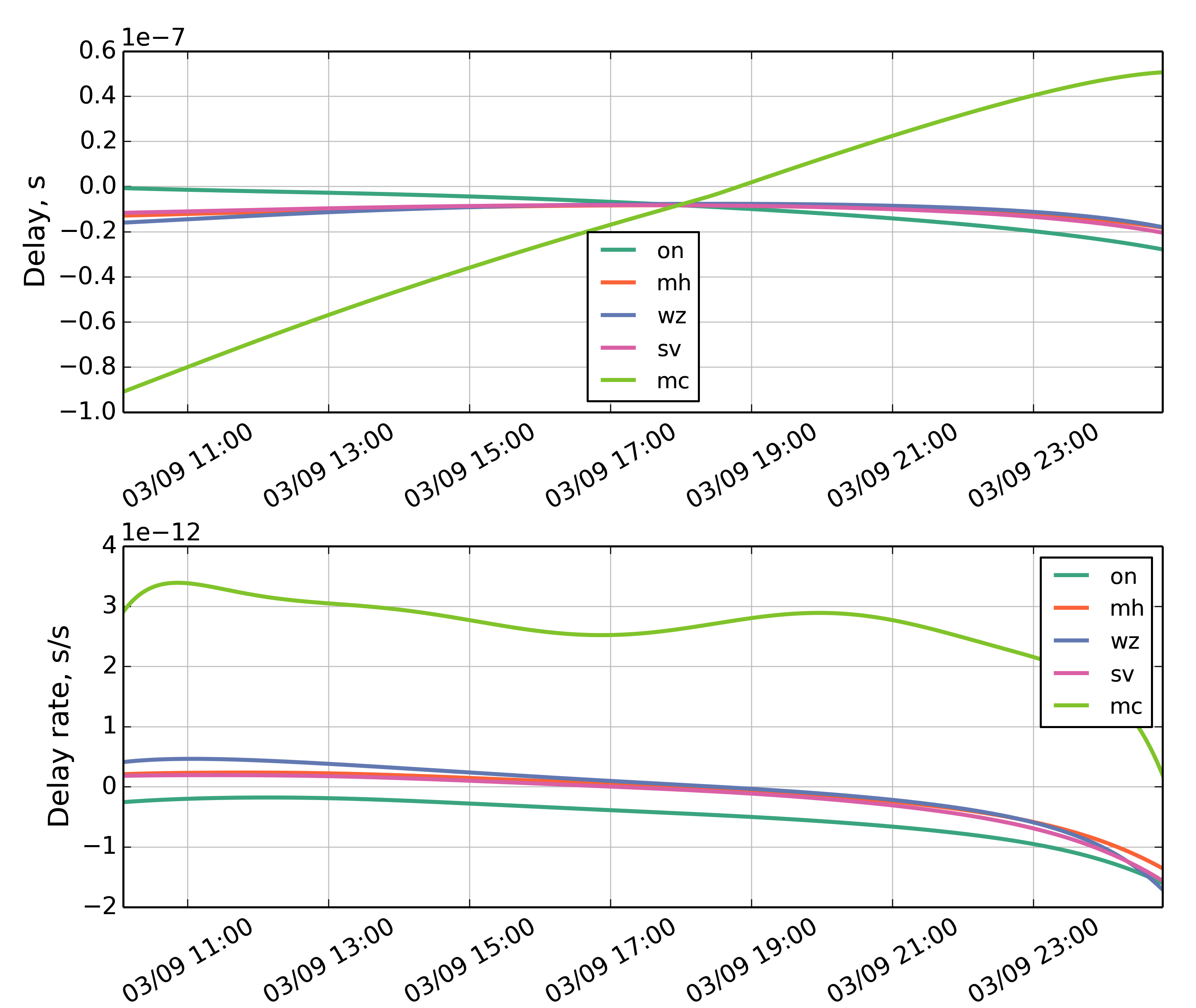}
   \caption{Signal delays ({\bf top}) and rates ({\bf bottom}) at stations due to propagation and instrumental effects. Experiment GK047B, 2013.03.09-10.}
   \label{DelRate}
   \end{figure}

For each station, the topocentric geometrical Doppler predictions $f_\mathrm{geom}(t)$ (shown by thin lines in Fig. \ref{OCNewOrbit}, top) were corrected for the instrumental effects (thermal deformation of the telescope dishes and axis offsets) and propagation in the Earth's troposphere and ionosphere to obtain the total Doppler predictions $f(t)$: 
\begin{equation}
f(t) = f_\mathrm{geom}(t -\tau_{ip} )\cdot(1-\dot{\tau}_{ip}),
\label{eq1}
\end{equation}
where $t$ is time in UTC, $\tau_{ip}$ and $\dot{\tau}_{ip}$ are the total signal delays and delay rates due to instrumental effects and propagation, respectively.

Equation \ref{eq1} immediately follows from the phase of an electromagnetic wave detected at a station $\varphi$ being related to the phase of the same wave in the absence of instrumental and propagation effects $\varphi_{geom}$ as $\varphi(t) = \varphi_{geom}(t-\tau_{ip})$ and from frequency being a time derivative of phase.

The delay rates at stations $\dot{\tau}_{ip}$ were computed numerically using the predictions of delays due to the mentioned effects (see Fig. \ref{DelRate})\footnote{Detailed description of the models used at this stage is given in \citet{2012A&A...541A..43D}.}. The results demonstrate the high accuracy of the models used in the PYP software (compare Fig. \ref{DelRate} (bottom) with Fig. \ref{OCNewOrbit} (middle) and the result of taking the effects into account in Fig. \ref{OCNewOrbit} (bottom)).

When comparing Fig. \ref{OCOldOrbit} (bottom) with Fig. \ref{OCNewOrbit}, an improvement in the orbital solution accuracy (two orders of magnitude for the velocity) is clearly seen. An offset of $\sim0.2$ Hz in differential frequency at 8.4 GHz corresponds to an error of $\sim0.7$ cm/s in velocity for the nominal orbit, whereas for the improved orbit it is $\sim1$ mHz corresponding to $\sim$35 $\mu$m/s (with $\sigma=3$ mHz, or 0.1 mm/s).

\begin{figure}
  \centering
  \includegraphics[width=260pt, clip]{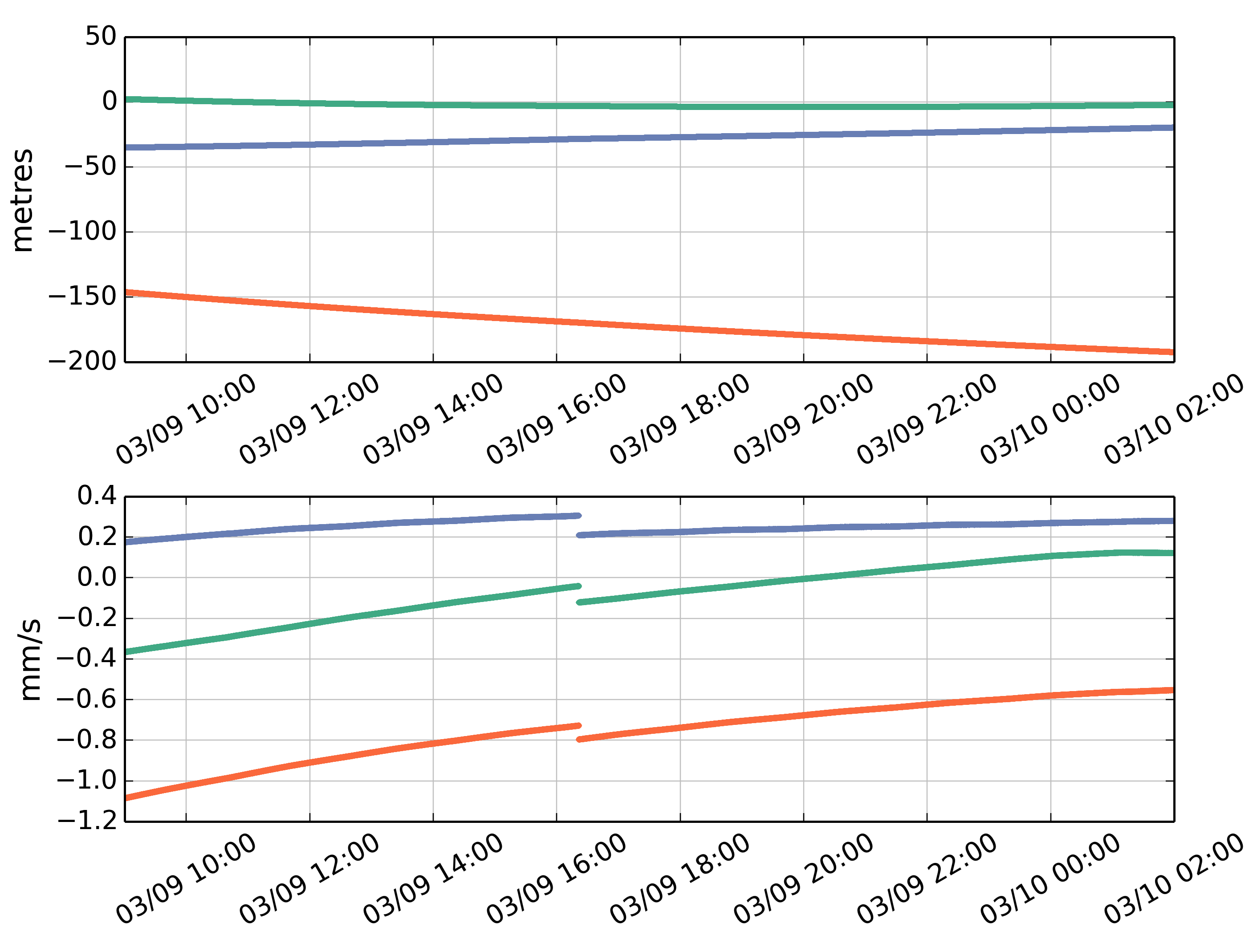}
  \caption{Difference between orbits obtained with and without PRIDE
    data. Position ({\bf top}) and velocity ({\bf bottom}) offsets are
    projected on the axis of orbital reference frame: radial -- green, normal
    -- orange, binormal~--~blue. Reaction wheel unloading was carried out around 16:20, hence the discontinuity in velocity difference.}
  \label{fig:orb_diff}
\end{figure}

\begin{figure}
  \centering
  \includegraphics[width=260 pt, clip]{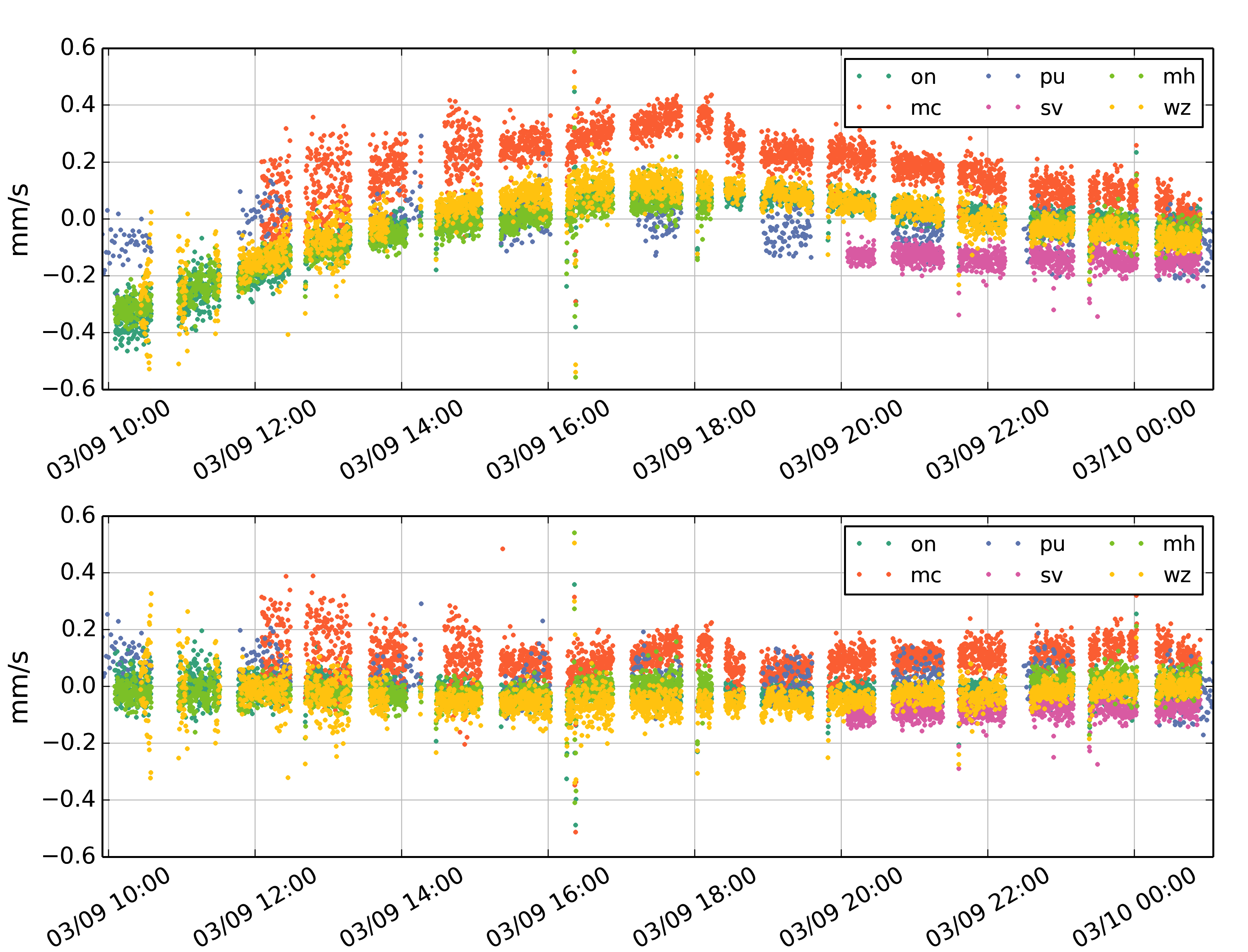}
  \caption{One-way Doppler (Pushchino + PRIDE observations) data fit (O$-$C)
    on the initial ({\bf top}) and the improved ({\bf bottom}) orbital
    solutions.}
  \label{fig:doppler_diff}
\end{figure}

The Doppler data gathered during these PRIDE-runs were fed into the orbit estimation pipeline described in Appendix \ref{app2}. 
Two orbital solutions were obtained: the initial one -- by using only the standard OD data and the improved ones by including the PRIDE Doppler observations into the OD process. Comparison of those orbits has shown a difference up to $\sim200$ m in position and slightly above 1 mm/s in velocity. According to the comparison and also consistent with covariance analysis, the most significant part of the difference is located in the normal direction, which lies in the orbital plane orthogonal to the radial direction (see Fig.~\ref{fig:orb_diff}). As expected, the covariance matrix of the improved orbit demonstrated a significant increase of accuracy of the velocity part of the spacecraft state vector during the experiment. Semi axes of the corresponding ellipsoid ($1\sigma$ level) have decreased from [0.70, 0.18, 0.39]~mm/s to [0.49, 0.13, 0.31]~mm/s. As for the formal estimation of positional accuracy, the use of PRIDE Doppler observations significantly improves the accuracy in the radial and binormal directions. However, error in the normal direction, which is the limiting factor for the spatial accuracy, has only decreased by 11\%. In contrast to the formal assessments, the accuracy of the improved orbit is clearly demonstrated by a comparison of the observed and calculated Doppler values on both orbits. As Fig.~\ref{fig:doppler_diff} (top) shows, the initial orbit contains systematic errors in velocity up to $\sim0.35$ mm/s and in acceleration up to $\sim1.1\cdot10^{-8}$ m/s$^2$. On the other hand, the improved orbit is almost free of such biases in the well-measured radial direction (see Fig.~\ref{fig:doppler_diff} (bottom)).

\subsection{Correlation of SVLBI data at JIVE.}

   \begin{figure}
   \centering
   \includegraphics[width=260pt,clip]{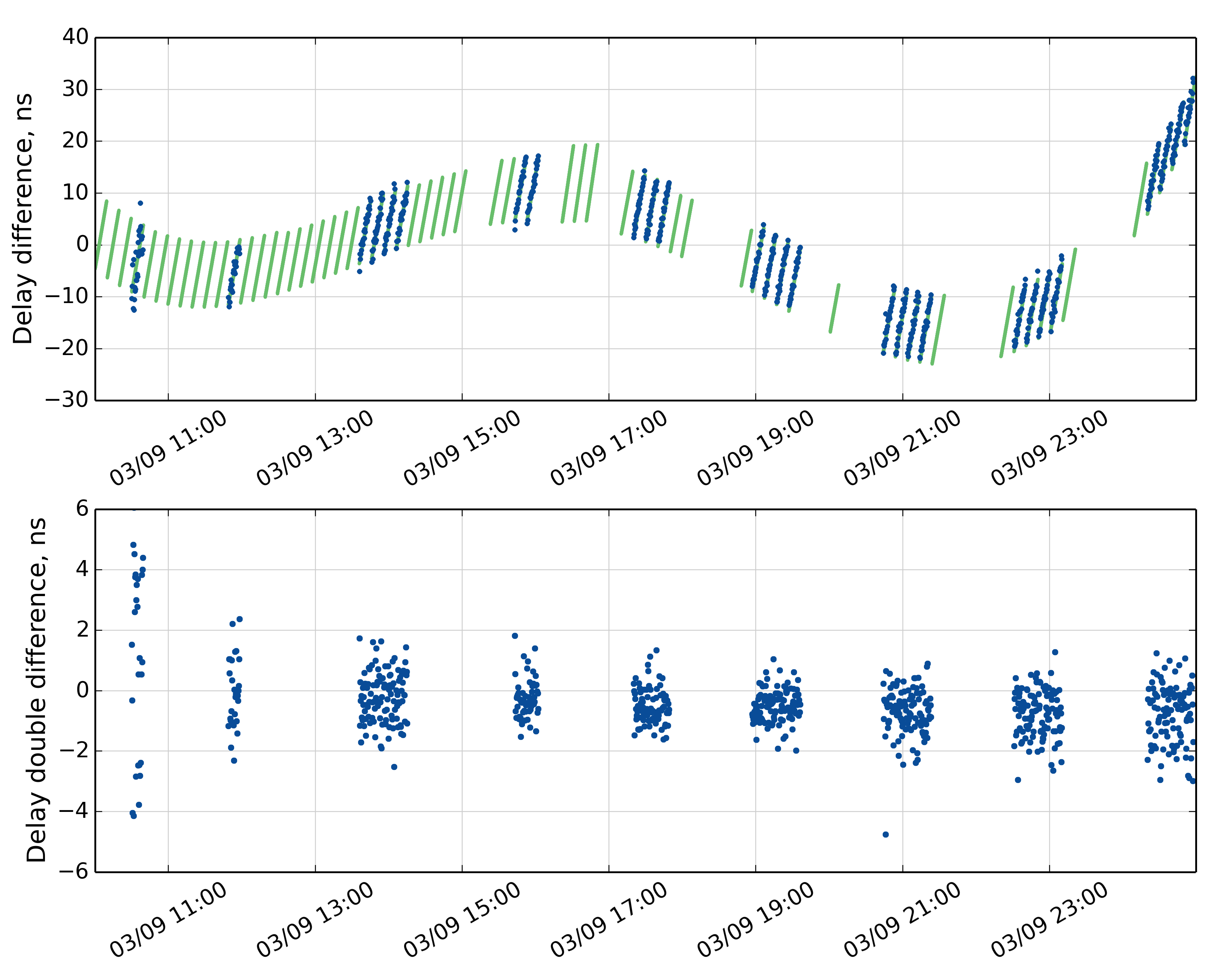}
   \caption{Improved versus nominal orbital solution: difference in measured residual delays (blue dots) compared to the difference in modelled delays (green dots) ({\bf top}) and their double difference ({\bf bottom}). Baseline Effelsberg -- RadioAstron. Experiment GK047A, 2013.03.09-10.}
   \label{FigDel}
    \end{figure}
%

The data of the experiment GK047A were correlated using the EVN Software Correlator at JIVE SFXC \citep{SFXC}. The VLBI delay model used for correlating RadioAstron data differs significantly from the one used in the standard Earth-based VLBI. First, the model from \citet{2012ARep...56..984V} is used instead of the conventional one from the IERS Technical Note 36 \citep{tn36}. This model \citep{2012ARep...56..984V} takes into account all terms up to $O((v/c)^3)$ order and includes all terms that depend on acceleration of both the Earth and the spacecraft. It provides a better-than-1-ps accuracy of delay $\widetilde{\Delta T}$ prediction for the case where one antenna is orbiting the Earth at a distance comparable to that of the Moon.

Second, as mentioned above, the RadioAstron data are not recorded on board, but rather streamed down by the 15 GHz link to a tracking station. Because of a hardware problem, time-stamp information in the data stream is lost \citep{James}. Therefore, the first frame of a scan gets locked to the local oscillator (LO) at the receiving station, after which the signal is treated simply as a byte stream. Thus the light travel time $LT_{SG}$ from the RadioAstron to the Ground receiving station has to be subtracted from the pre-calculated delays for RadioAstron in the TT time scale\footnote{The terrestrial time TT is the theoretical time--scale for clocks at sea-level. It differs by 32.184 seconds from the international atomic time scale TAI, the only physically realised time--scale. The latter, in turn, differs by a number of leap-seconds from the time--scale used for timing measurements at stations -- the coordinated universal time UTC. There is only an offset between these time scales, but no rate.} for each of the scan start times. Instrumental and propagation delays are included in $LT_{SG}$. We note that the ionospheric delay here is calculated for the frequency of 15 GHz, because RadioAstron down-streams data at this frequency.

Third, one needs to account for the gravitational redshift \citep{James}. In the framework of the IAU resolutions \citep{tn36}, the proper time $\tau_{S}$ of a spacecraft located at the position ${\bf x}_{S}$ and having the coordinate velocity ${\bf v}_{S}$ in the geocentric celestial reference system (GCRS) can be transformed into the TT time scale


\begin{eqnarray}
\frac{\mathrm{d}\tau_{S}}{\mathrm{d}TT} = 1 + L_G - 
  \frac{1}{c^2}\bigg[ \frac{{\bf v}_{S}^{2}}{2} + U_E({\bf x}_{S}) + V({\bf X}_{S}) - \nonumber\\
    V({\bf X}_{E}) - x_{S}^i \partial_i V({\bf X}_{E}) \bigg] = 1 + L_G + f(TT),
\label{dtaudtt}
\end{eqnarray}
where for brevity we introduced the function
\begin{equation}
f(TT) = - 
  \frac{1}{c^2}\bigg[ \frac{{\bf v}_{S}^{2}}{2} + U_E({\bf x}_{S}) + V({\bf X}_{S}) - 
    V({\bf X}_{E}) - x_{S}^i \partial_i V({\bf X}_{E}) \bigg].
\end{equation}
Here, $L_G = 6.969290134 \times 10^{-10}$, $t_0$ is the start epoch of a scan, $U_E$ is the Newtonian potential of the Earth at ${\bf x}_{S}$ in the geocentric system GCRS, $V$ the sum of Newtonian potentials of the other bodies (mainly the Sun and the Moon), either at the Earth's centre of mass ${\bf X}_E$ or at the spacecraft location ${\bf X}_S$ in the barycentric celestial reference system BCRS:

\begin{equation}
U_E({\bf x}_{S}) = \frac{GM_E}{|{\bf x}_{S}|} + \frac{G M_E R_E^2 J_2}{2|{\bf x}_{S}|^3}
  \bigg( 1 - \frac{3z_{S}^2}{|{\bf x}_{S}|^2} \bigg),
\label{UE}
\end{equation}

\begin{equation}
V({\bf X}_{E}) = \sum_J \frac{GM_J}{|{\bf R}_{\oplus J}|}, \hspace{10pt}
V({\bf X}_{S}) = \sum_J \frac{GM_J}{|{\bf R}_{S J}|},
\label{VE}
\end{equation}

\begin{equation}
 x_{S}^i \partial_i V({\bf X}_{E}) = - \sum_J GM_J \frac{{\bf x}_{S} \cdot {\bf R}_{\oplus J}}{|{\bf R}_{\oplus J}|^3}.
\label{lastterm}
\end{equation}
Here ${\bf R}_{\oplus J}$ and ${\bf R}_{S J}$ are the vectors from the J-th gravitating body to the geocentre and to the spacecraft, respectively:

\begin{eqnarray}
{\bf R}_{\oplus J} = {\bf R}_{J}^{BCRS} - {\bf R}_{\oplus}^{BCRS} \approx {\bf r}_{J}^{GCRS}
\label{REJ} \\
{\bf R}_{SJ} = {\bf R}_{J}^{BCRS} - {\bf R}_{S}^{BCRS} \approx {\bf r}_{J}^{GCRS} - {\bf r}_{S}^{GCRS}
\label{RSJ}.
\end{eqnarray}


To account for the redshift, one needs to integrate Eq. \ref{dtaudtt} over a scan span time, which yields

\begin{equation}
\tau_{S} \bigg|_t= (1 + L_G) \cdot (t - t_0) + \int_\mathrm{t_0}^{t} f(TT) \mathrm{d}TT
\label{tautt}.
\end{equation}

The final formula for the delay $\Delta T$ in the TT frame to be fed into the correlator is given by the following equation:

\begin{equation}
\Delta T = \widetilde{\Delta T} + LT_{SG} + \tau_{S}.
\end{equation}

The fringes on ground-space baselines were detected using the delay model for RadioAstron described above. Both delay sets calculated using the nominal and the improved orbital solutions yielded strong fringes. The results of the cross-correlation were fringe-fitted using the AIPS software \citep{2003ASSL..285..109G}, as well as the software tools, developed at JIVE. The residual clock offsets and clock rates, as free parameters, were removed from both of the solutions. The difference in measured residual delays was compared to the difference in modelled delays. The results of this comparison are shown in Fig. \ref{FigDel} (top) for the baseline Effelsberg -- RadioAstron. The observed and the computed values show good agreement demonstrated by the double difference, which is presented in Fig. \ref{FigDel} (bottom).

Although in this particular case the improved orbit is not critical, it will be useful for weaker sources, when a considerable `clock acceleration' does not allow using longer integration times to detect fringes.

\section{Discussion and conclusions}
We have presented a new approach to orbit determination of the RadioAstron spacecraft. When applied utilising multi-station Doppler tracking data, it provided a significant improvement in the accuracy of the orbital solution, which was demonstrated and validated using both Doppler and VLBI measurements.
However, we need to point out that the positional accuracy of RadioAstron can be improved further if one includes the VLBI observations of the RadioAstron downlink signal (in addition to Doppler and other measurements used currently) in the orbit determination process. Several technical problems have to be solved in that concern, with the first one being continuous tracking of spacecraft by ground telescopes. The spacecraft's angular motion on the celestial sphere is quite fast and can even be anti-sidereal. Therefore, if tracking is not continuous, i.e. the telescopes are repointed every few seconds\footnote{This is constrained by the telescope motor's characteristics. Starting and stopping them too often is not technically possible.}, the spacecraft will constantly move out of the primary (synthesised) beam of the interferometric array. Although the spacecraft signal will still be detected via the side lobes, this will result in `glitches' when deriving the group delay of the observed signal at times when the telescopes are repointed. These can reach several nanoseconds in magnitude, practically eliminating the benefit of using such data.

\begin{figure}
   \centering
   \includegraphics[width=260pt,clip]{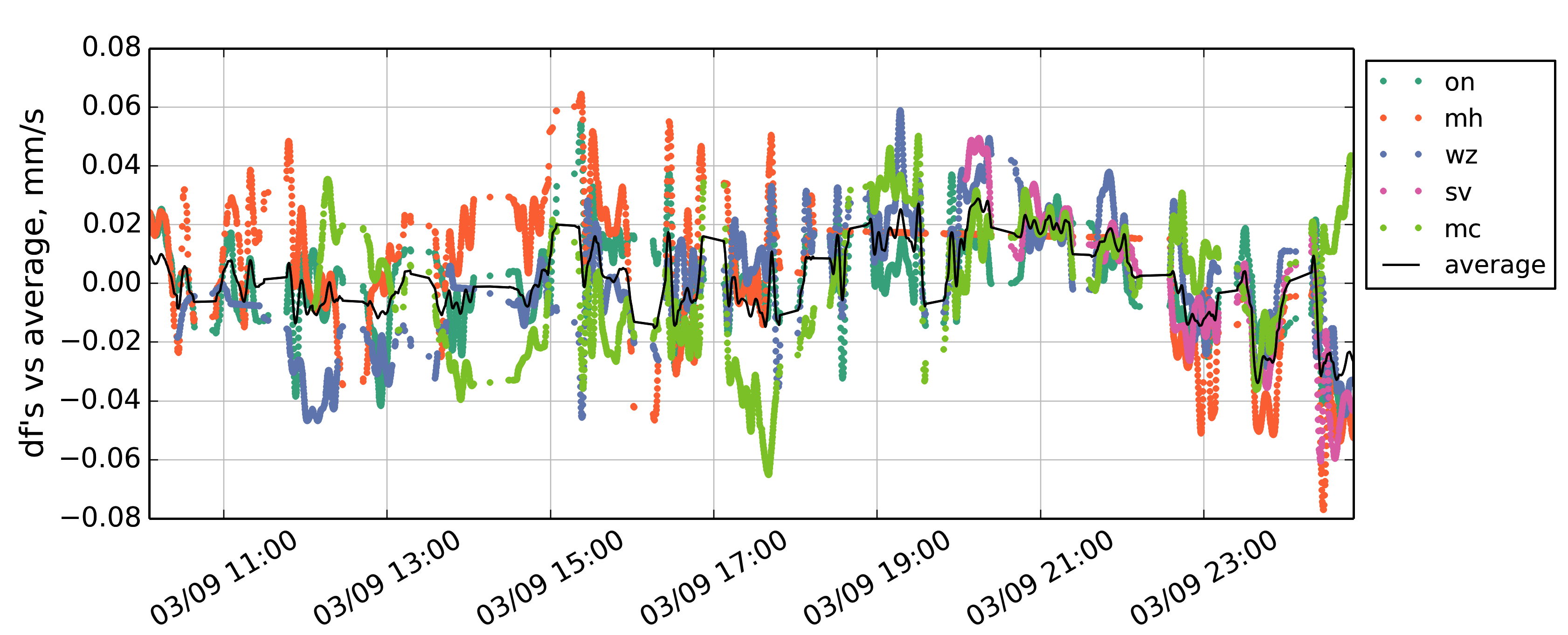}
   \caption{Residual Doppler noise. Experiment GK047B.}
   \label{DopNoise}
   \end{figure}

Further improvement of the state vector determination accuracy, in addition to scientific applications of the RadioAstron mission mentioned in the introduction, will be beneficial for an initiative to measure the gravitational redshift of frequency $\Delta f$ experienced by an electromagnetic wave travelling in a region of space with varying gravitational potential \citep{1973grav.book.....M}. Precise measurement of this quantity will make it possible to achieve a better constraint on the value of the post-Newtonian parameter $\gamma$, which is equal to unity in general relativity, but which deviates from it slightly in alternative models. Accurate Doppler measurements of the RadioAstron's downlink signal might play an important role in such an experiment (for details, see \citet{Mitya}). To demonstrate the achievable precision, we performed the following test. The filtered differential frequency time series from Fig. \ref{OCNewOrbit} (bottom) for each station were smoothed to suppress short-term variations, then the mean values were subtracted from the data and a weighted average over all stations was computed. The residual Doppler noise is on average at the level of 10-20 $\mu$m/s in 250 seconds (see Fig. \ref{DopNoise}). This sets a limit for the accuracy of experimental determination of $\gamma$ at the level of $1.8\cdot10^{-5}$ over $\sim1000$ hours if no noise-compensation scheme is used. In case such a scheme is utilised, the accuracy can be improved by an order of magnitude \citep{Mitya}, which is comparable to or even better than the result obtained using Cassini spacecraft tracking during solar conjunction \citep{2003Natur.425..374B}.

Another important issue in this context is the accuracy of clock offset time series provided by the stations. Usually, these are given with a time resolution of one day with a typical rounding error of $\sim 1$ ns. This is sufficient for using these data for their main purpose -- as a priori values when searching for the clock parameters (offset and rate) of stations with respect to a reference station when performing the VLBI correlation. However, such accuracy can result in a significant error in the calculated delay rates used in the Doppler analysis, where absolute values are used. For instance, the differential frequency offset for station Medicina clearly seen in Fig. \ref{OCNewOrbit} (bottom, green dots), is at the level of 3 mHz, which corresponds to a delay rate of 0.3 ps/s, which in turn could be easily explained by an insufficient accuracy of the time series of the nominal H-maser versus GPS clock offsets. Therefore, we suggest that stations keep clock offset time series with a denser time resolution and with more digits in the readings.

%
%


\begin{acknowledgements}
      We would like to express our gratitude to the referee M.~Reid (Harvard-Smithsonian Center for Astrophysics) for useful and constructive suggestions. The EVN is a joint facility of European, Chinese, South African, and other radio astronomy institutes funded by their national research councils. The authors would like to thank the personnel of the participating stations: P.~de Vicente (Yebes), J.~Quick (Hartebeesthoek), G.~Kronschnabl (Wettzell), R.~Haas (Onsala), A.~Orlatti (Medicina), A.~Ipatov, M.~Kharinov, and A.~Mikhailov (Svetloe of the KVAZAR network). We would also like to thank K.V.~Sokolovsky and Y.Y.~Kovalev (ASC RAS) for assistance. R.M.~Campbell, A.~Keimpema, and P.~Boven of JIVE provided important support to various components of the project. G.~Cim\'{o} acknowledges the EC FP7 project ESPaCE (grant agreement 263466). T.~Bocanegra Bahamon acknowledges the NWO--ShAO agreement on collaboration in VLBI.
\end{acknowledgements}

\bibliographystyle{aa} 
\bibliography{ref} 

\begin{thebibliography}{17}
\expandafter\ifx\csname natexlab\endcsname\relax\def\natexlab#1{#1}\fi

\bibitem[{{Anderson}(2012)}]{James}
{Anderson}, J.~M. 2012, in DiFX Annual Meeting. Held 24 September 2012 in
  Sydney, Australia

\bibitem[{{Bertotti} {et~al.}(2003){Bertotti}, {Iess}, \&
  {Tortora}}]{2003Natur.425..374B}
{Bertotti}, B., {Iess}, L., \& {Tortora}, P. 2003, \nat, 425, 374

\bibitem[{{Biriukov} {et~al.}(2014){Biriukov}, {Kauts}, {Kulagin}, {Litvinov},
  \& {Rudenko}}]{Mitya}
{Biriukov}, A.~V., {Kauts}, V.~L., {Kulagin}, V.~V., {Litvinov}, D.~A., \&
  {Rudenko}, V.~N. 2014, Astronomy Reports, 58, 783

\bibitem[{{Duev}(2012)}]{duevPhD}
{Duev}, D.~A. 2012, PhD thesis, {Lomonosov Moscow State University}

\bibitem[{{Duev} {et~al.}(2012){Duev}, {Molera Calv{\'e}s}, {Pogrebenko},
  {Gurvits}, {Cim{\'o}}, \& {Bocanegra Bahamon}}]{2012A&A...541A..43D}
{Duev}, D.~A., {Molera Calv{\'e}s}, G., {Pogrebenko}, S.~V., {et~al.} 2012,
  \aap, 541, A43

\bibitem[{{Greisen}(2003)}]{2003ASSL..285..109G}
{Greisen}, E.~W. 2003, Information Handling in Astronomy - Historical Vistas,
  285, 109

\bibitem[{{Gurvits}(1991)}]{Gurvits91}
{Gurvits}, L.~I. 1991, in Frontiers of VLBI, eds. H. Hirabayashi, M. Inoue, H.
  Kobayashi, Universal Academy Press, 187

\bibitem[{{Kardashev} {et~al.}(2013){Kardashev}, {Khartov}, {Abramov},
  {Avdeev}, {Alakoz}, {Aleksandrov}, {Ananthakrishnan}, {Andreyanov},
  {Andrianov}, {Antonov}, {Artyukhov}, {Arkhipov}, {Baan}, {Babakin},
  {Babyshkin}, {Bartel'}, {Belousov}, {Belyaev}, {Berulis}, {Burke},
  {Biryukov}, {Bubnov}, {Burgin}, {Busca}, {Bykadorov}, {Bychkova},
  {Vasil'kov}, {Wellington}, {Vinogradov}, {Wietfeldt}, {Voitsik},
  {Gvamichava}, {Girin}, {Gurvits}, {Dagkesamanskii}, {D'Addario},
  {Giovannini}, {Jauncey}, {Dewdney}, {D'yakov}, {Zharov}, {Zhuravlev},
  {Zaslavskii}, {Zakhvatkin}, {Zinov'ev}, {Ilinen}, {Ipatov}, {Kanevskii},
  {Knorin}, {Casse}, {Kellermann}, {Kovalev}, {Kovalev}, {Kovalenko}, {Kogan},
  {Komaev}, {Konovalenko}, {Kopelyanskii}, {Korneev}, {Kostenko}, {Kotik},
  {Kreisman}, {Kukushkin}, {Kulishenko}, {Cooper}, {Kut'kin}, {Cannon},
  {Larionov}, {Lisakov}, {Litvinenko}, {Likhachev}, {Likhacheva}, {Lobanov},
  {Logvinenko}, {Langston}, {McCracken}, {Medvedev}, {Melekhin}, {Menderov},
  {Murphy}, {Mizyakina}, {Mozgovoi}, {Nikolaev}, {Novikov}, {Novikov},
  {Oreshko}, {Pavlenko}, {Pashchenko}, {Ponomarev}, {Popov}, {Pravin-Kumar},
  {Preston}, {Pyshnov}, {Rakhimov}, {Rozhkov}, {Romney}, {Rocha}, {Rudakov},
  {R{\"a}is{\"a}nen}, {Sazankov}, {Sakharov}, {Semenov}, {Serebrennikov},
  {Schilizzi}, {Skulachev}, {Slysh}, {Smirnov}, {Smith}, {Soglasnov},
  {Sokolovskii}, {Sondaar}, {Stepan'yants}, {Turygin}, {Turygin}, {Tuchin},
  {Urpo}, {Fedorchuk}, {Finkel'shtein}, {Fomalont}, {Fejes}, {Fomina},
  {Khapin}, {Tsarevskii}, {Zensus}, {Chuprikov}, {Shatskaya}, {Shapirovskaya},
  {Sheikhet}, {Shirshakov}, {Schmidt}, {Shnyreva}, {Shpilevskii}, {Ekers}, \&
  {Yakimov}}]{2013ARep...57..153K}
{Kardashev}, N.~S., {Khartov}, V.~V., {Abramov}, V.~V., {et~al.} 2013,
  Astronomy Reports, 57, 153

\bibitem[{{Keimpema} {et~al.}(2014){Keimpema}, {Kettenis}, {Pogrebenko}, M.,
  {Cim{\`o}}, {Duev}, {Eldering}, {Kruithof}, {van Langevelde}, {Marchal},
  {Molera Calv{\'e}s}, {Ozdemir}, {Paragi}, {Pidopryhora}, {Szomoru}, \&
  {Yang}}]{SFXC}
{Keimpema}, A., {Kettenis}, M.~M., {Pogrebenko}, S.~V., {et~al.} 2014,
  Experimental Astronomy, (to appear)

\bibitem[{{McCarthy} \& {Petit}(2004)}]{tn32}
{McCarthy}, D.~D. \& {Petit}, G., eds. 2004, {IERS Technical Note \# 32}

\bibitem[{{Misner} {et~al.}(1973){Misner}, {Thorne}, \&
  {Wheeler}}]{1973grav.book.....M}
{Misner}, C.~W., {Thorne}, K.~S., \& {Wheeler}, J.~A. 1973, {Gravitation}

\bibitem[{{Molera Calv\'{e}s}(2012)}]{GuifrePhD}
{Molera Calv\'{e}s}, G. 2012, PhD thesis, {Aalto University}, {Pub. No.
  42/2012}

\bibitem[{{Molera Calv{\'e}s} {et~al.}(2014){Molera Calv{\'e}s}, {Pogrebenko},
  {Cim{\`o}}, {Duev}, {Bocanegra-Baham{\'o}n}, {Wagner}, {Kallunki}, {de
  Vicente}, {Kronschnabl}, {Haas}, {Quick}, {Maccaferri}, {Colucci}, {Wang},
  {Yang}, \& {Hao}}]{2014A&A...564A...4M}
{Molera Calv{\'e}s}, G., {Pogrebenko}, S.~V., {Cim{\`o}}, G., {et~al.} 2014,
  \aap, 564, A4

\bibitem[{{Motenbruck} \& {Gill}(2001)}]{MotenbruckGill}
{Motenbruck}, O. \& {Gill}, E. 2001, {Satellite Orbits: Models, Methods, and
  Applications}

\bibitem[{{Petit} \& {Luzum}(2010)}]{tn36}
{Petit}, G. \& {Luzum}, B., eds. 2010, {IERS Technical Note \# 36}

\bibitem[{{Standish}(1998)}]{Standish}
{Standish}, E.~M., ed. 1998, {JPL Planetary and Lunar Ephemerides DE405/LE405.}

\bibitem[{{Vlasov} {et~al.}(2012){Vlasov}, {Zharov}, \&
  {Sazhin}}]{2012ARep...56..984V}
{Vlasov}, I.~Y., {Zharov}, V.~E., \& {Sazhin}, M.~V. 2012, Astronomy Reports,
  56, 984

\end{thebibliography}


\begin{appendix}
\section{Dynamical model of RadioAstron motion}\label{app1}
\begin{table*}
  \caption[Description of the RadioAstron dynamical model]{Description of the RadioAstron dynamical model\footnotemark}
  \centering
  \begin{tabular}{ll}
    \hline
    \hline\\
    Gravity      & \\
    \hline
    Static             & EGM96 up to 75th degree/order \\
    Third bodies       & DE405 Sun, Moon, and all planets \\
    Earth tides        & IERS 2003 conventions \\
    Pole tides         & Solid Earth \\
    Ocean tides        & FES2004 up to 20th degree/order  \\
    General relativity & IERS 2010 conventions \\
                 & \\
    Surface forces   & \\
    \hline
    Solar radiation  & Box-wing + SRT (3 parameters) \\
    Earth radiation  & Applied (18x9 elements) \\
    Atmospheric drag & GOST R 25645.166-2004 density model\\
                 & \\
    Reaction wheels unloadings \\
    \hline
    Direction          & Spacecraft attitude from on-board star sensors and
                         attitude event file \\
    Delta V value      & From known mass of spent propellant and a priori thrust values  \\
                       & Measurements of angular momentum change of reaction wheels \\
    \hline
  \end{tabular}
  \label{tab:dyn_mod}
\end{table*}

\begin{figure}
  \centering
  \includegraphics[width=120pt]{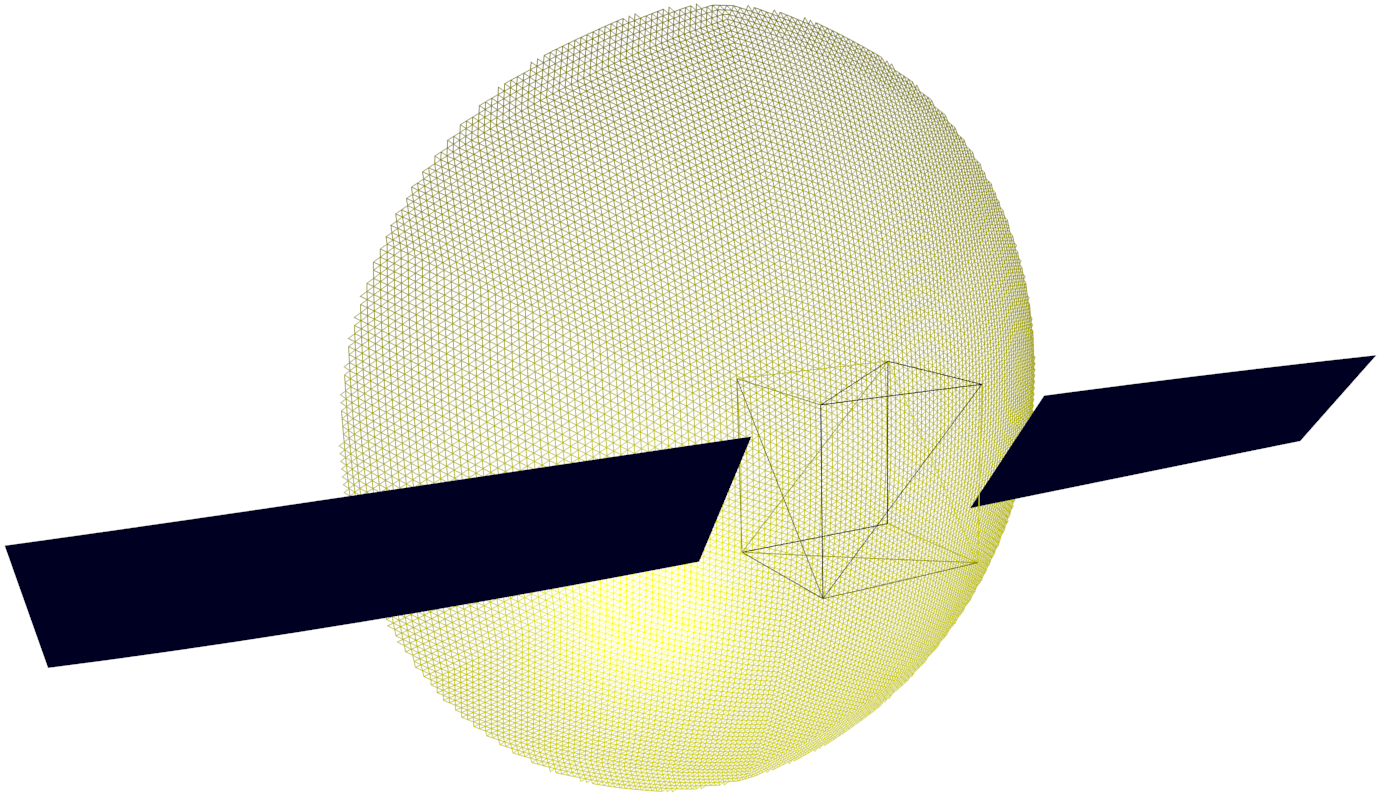}
  \includegraphics[width=120pt]{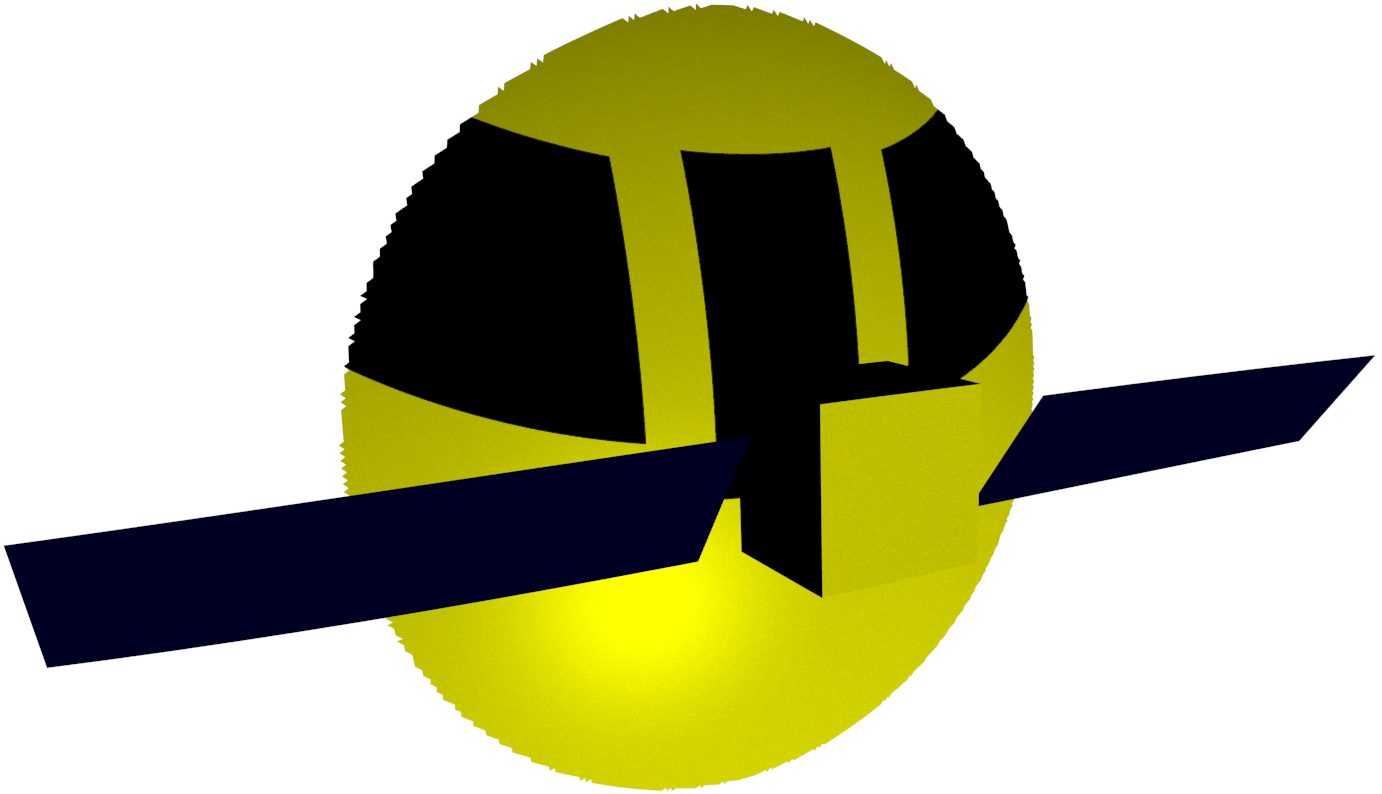}
  \caption{Simplified model of the spacecraft surface: SRT antenna, spacecraft bus, solar panels (left panel). Permissible attitude of the spacecraft with respect to the Sun prevents the sunlight from getting to the antenna \citep{Gurvits91}. Therefore, a complex shadowing occurs only on the SRT antenna surface (right panel; the spacecraft is illuminated from the bottom right corner).}
  \label{fig:ra_surf}
\end{figure}

The motion of the centre of mass of RadioAstron is considered passive between adjacent unloadings. The force model, which was used in corresponding equations of motion on passive arcs, is reflected in Table \ref{tab:dyn_mod}. We only go into detail about the SRP model, since the models of other perturbations are fairly common (see, e.g., \citealt{MotenbruckGill}). The SRP model is based on the analytical calculation of solar flux pressure, which is divided into absorbed, reflected specularly and reflected diffusely parts. Net SRP is calculated by integration over a simplified spacecraft surface. The model surface consists of the parabolic antenna of the space radio telescope (SRT), spacecraft bus, and solar panels (see Fig. \ref{fig:ra_surf}). As in a so-called box-wing model, the surfaces of spacecraft bus and solar panels are flat rectangles. The SRT surface is split into multiple flat elements in order to take the shadowing from other parts of the spacecraft into account.

Reflectivity and specularity of the corresponding surfaces are used as the parameters of the model. Since the spacecraft bus and the illuminated parts of SRT are covered with a multi-layer insulation (MLI), the optical properties of these surfaces were believed to be the same. For solar panels, in order to avoid correlation between the parameters, the specularity value was set to unity. Thus the SRP model is adjusted by means of three parameters: specularity $\mu_1$ and reflectivity $\alpha_1$ of MLI and reflectivity $\alpha_2$ of the solar panels. It is noteworthy that the model also allows estimating perturbing SRP torque, as far as the simplified surface only consists of flat elements with known coordinates of centres of pressure with respect to the centre of mass.

The effect of an unloading of reaction wheels on the spacecraft motion is incorporated into the dynamical model because the instantaneous velocity increment $\Delta\mathbf{v}$ occurred at a known time. The whole process of unloading of reaction wheels takes less than a couple of minutes and consists of several firings of stabilisation thrusters. The time of $\Delta\mathbf{v}$ application is calculated as a weighted average of the times of firings during the unloading. In this approach, the impact of unloadings can be described with a set of pairs $\{t_i, \Delta\mathbf{v}_i\}_{i=1}^n$ of application times and velocity increments. While times $t_i$ are considered to be known accurately enough, the corresponding $\Delta\mathbf{v}_i$ are assumed to be the parameters of the dynamical model to be refined in the course of the OD process.


\footnotetext{Detailed description of the models can be found in \citealt{Standish, tn32, tn36}, and at  \url{http://www.gosthelp.ru/gost/gost5827.html}}

\section{Orbit estimation pipeline}\label{app2}

The RadioAstron orbit determination process uses a least-squares algorithm, in which parameters of dynamical and measurement models are iteratively adjusted to provide the best fit of the given tracking data. The dynamical parameters comprise the initial state vector (position and velocity) of the spacecraft, parameters of the SRP model and velocity increments $\{\Delta\mathbf{v}_i\}_{i=1}^n$ due to $i-th$ unloading of reaction wheels (with a total of $n$ unloadings for one orbital arc). Among the parameters of the measurement model are per-pass range biases and a constant spacecraft clock offset rate that is causing a shift in one-way Doppler observables. These parameters are determined using orbital arcs of up to 50 days long.

On-board measurements provide additional estimates of the parameters of unloadings $\{\Delta\mathbf{v}_i^0\}_{i=1}^n$ using attitude data, thrusters' operation parameters, and the measured values of the unloaded angular momentum of reaction wheels. We tried to keep the dynamical parameters of the model $\{\Delta\mathbf{v}_i\}_{i=1}^n$ as close as possible to those estimates by adding the following term to the functional to be minimised
\begin{equation}
  \label{eq:func_dv}
  \Delta\Phi_{RW} = \sum_{i=1}^n(\Delta\mathbf{v}_i^0 -
  \Delta\mathbf{v}_i)^{\mathsf{T}}\cdot
  \mathbf{P}_i\cdot(\Delta\mathbf{v}_i^0 -
  \Delta\mathbf{v}_i),
\end{equation}
where $\mathbf{P}_i$ is a weight matrix for $\Delta\mathbf{v}_i^0$. The matrix is calculated under an assumption that the direction of the estimated vector is determined with rather small errors. It is true in the case of RadioAstron since the spacecraft holds its attitude during unloading. The weight matrix is therefore given as follows
\begin{equation}
  \label{eq:imp_weight}
  \mathbf{P}_i = \sigma_d^{-2}(\mathbf{E} - \mathbf{e}_i\cdot\mathbf{e}_i^{\mathsf{T}}) + 
  \sigma_v^{-2}\mathbf{e}_i\cdot\mathbf{e}_i^{\mathsf{T}},
\end{equation}
where $\mathbf{e}_i$ is the nominal direction of $\Delta\mathbf{v}_i$, $\sigma_v$ and $\sigma_d$ are the corresponding errors along $\mathbf{e}_i$ and orthogonal to it (magnitude and direction errors), $\mathbf{E}$ is the identity matrix. Values for determining the weight matrix deviations are selected to correspond to a 3\% error of $|\Delta\mathbf{v}_i|$ magnitude estimation and a 0.5 degree error in direction estimate.

Measurements of the rotation speeds of reaction wheels, hence the angular momentum of the system, allow us to not only verify $\Delta\mathbf{v}_i$ due to unloadings, but also to estimate the external torque, which is mostly caused by SRP, and to calibrate the SRP model. Torque measurements are performed, while the spacecraft is at a constant attitude, such as during observations and on altitudes higher than $10^5$ km in order to diminish the effect of Earth-related torques, like gravitational torque. These measurements are used along with the standard tracking data in the OD process with the following residuals, whose weighted squares are making the corresponding addition to the functional to be minimised:
\begin{equation}
  \label{eq:torque_meas}
  \xi = \frac{\mathbf{L}(t_2) - \mathbf{L}(t_1)}{t_2-t_1} - \mathbf{M}_{SRP}(R_{\odot},\Lambda,
  \alpha_1, \mu_1, \alpha_2).
\end{equation}
Equation \eqref{eq:torque_meas} is written in a spacecraft fixed reference frame, $\mathbf{L}(t)$ is the measured angular momentum of the reaction wheels, $(t_1,t_2)$ a timespan of an observation, $\mathbf{M}$ the computed SRP torque depending on the Sun distance $R_{\odot}$, spacecraft attitude $\Lambda$, and parameters of the model $\alpha_1$ , $\mu_1$ and $\alpha_2$.
Additional data provided by torque measurements contribute to the accuracy of the SRP model and also significantly reduce the correlation between the parameters of the model.

\end{appendix}

\end{document}